\documentclass[preprint,showpacs,aps,pre,amsmath,amssymb]{revtex4}

\usepackage{amsmath}
\usepackage{graphicx}
\usepackage{dcolumn}
\usepackage{bm}

\begin{document}
\title{Volume fluctuations and linear response in a simple model of compaction}
\author{J. Javier Brey}
\author{A. Prados}
\affiliation{F\'{\i}sica Te\'{o}rica, Universidad de Sevilla,
Apartado de Correos 1065, E-41080, Sevilla, Spain}

\date{\today }

\begin{abstract}
By means of a simple model system, the total volume fluctuations of
a tapped granular material in the steady state are studied.  In the
limit of a system with a large number of particles, they are found
to be Gaussian distributed, and  explicit expressions for the
average and the variance are provided. Experimental and molecular
dynamics results are analyzed and qualitatively compared with the
model predictions. The relevance of considering open or closed
systems is discussed, as well as the meaning and properties of the
Edwards compactivity and the effective (configurational) temperature
introduced by some authors. Finally, the linear response to a change
in the vibration intensity is also investigated. A KWW decay of the
volume response function is clearly identified. This seems to
confirm some kind of similarity between externally excited granular
systems and structural glasses.

\end{abstract}

\pacs{47.50.Cc, 05.50.+q, 81.05.Rm}

\maketitle

\section{Introduction}
\label{s1} Compaction can be roughly defined as the density
relaxation of a loosely packed system of many grains under
mechanical tapping or vibration \cite{ByM93,KFLJyN95}. Both the time
relaxation of the system and the steady state eventually reached in
the long time limit, exhibit interesting properties, and have
received a lot of attention during the last years. The description
of this phenomenon goes well beyond the scope of equilibrium
statistical physics and even of usual non-equilibrium theories. The
time evolution of the system corresponds to a consecutive series of
mechanically stable states in which all the particles are at rest,
at least at a macroscopic level of description. Therefore, the
system is not thermal, in the sense that the fluctuations of the
particles around their average positions play no role in
characterizing the state of the system. In fact, in the description
of the phenomenology of compaction, time is usually measured in
number of taps that, of course, has nothing to do with the real
motion of the particles.

Edwards and collaborators \cite{EyO89} formulated some years ago a
theory providing a thermodynamic-like description of granular
systems under certain circumstances. The main hypothesis of the
theory is that for externally perturbed powders in a steady state,
all the mechanically stable (metastable) configurations of a
granular assembly occupying the same volume have the same
probability. Therefore, it is assumed that the volume plays here a
role analogous to that of the energy in usual thermal systems. The
parameter conjugated to the volume, similar to the thermal
temperature, was named `compactivity'.

In principle, the conditions under which the theory should apply are
fulfilled in the compaction process. For this reason, the
predictions of the Edwards theory have been compared with results
obtained in experiments \cite{NKBJyN98,PhyB02,SGyS05}, numerical
simulations \cite{MyK02,CCyN06}, and simple models
\cite{BKLyS00,FNyC02,DyL03,BPyS00,LyD01,BFyS02,SGyL02,PyB02,ByP03a}
of compaction. Although in general a fairly good agreement is
claimed, at least in the limit of weak tapping, an analysis of the
situation shows that there are many fundamental aspects that can not
be considered as reasonably well established. A few of them will be
addressed, although not solved with generality, here:
\begin{enumerate}
\item Given that many of the experimental determinations of the
compactivity are based in the measurements of density fluctuations,
special care must be given to differentiate between fluctuations in
the number of particles in a system of constant volume, fluctuations
of volume in a system with a given number of particles, and
fluctuations of density in a system whose volume and number of
particle change in time. All these fluctuations are not equivalent,
in principle \cite{ByP03a}.

\item Some authors have reformulated the thermodynamic theory of
powders using the (configurational) energy as the variable
characterizing the macroscopic state of the system and an {\em
effective} temperature, i.e. they use a formulation much closer to
ordinary statistical physics. The equivalence or relationship
between both formulations does not seem to be clear, although some
ideas have been presented \cite{CyN01,MyK02,CCyN06}. For instance,
are the compactivity and the effective temperature independent
variables or there is a relationship between them?

\item The thermodynamic theory of powders was formulated for complete systems,
and not as a local theory. In this context, it could be thought that
a column of tapped granular matter in the steady state is
characterized by a unique compactivity, aside from the boundary
layers, even though the system is inhomogeneous. Another possibility
is to consider that the system can be divided into horizontal
layers, each of them having a different compactivity (or effective
temperature). In fact, this local version of the theory has been
often used in describing experiments and particle simulations. Which
are the physical meaning and the implications of a gradient of the
compactivity, or effective temperature, are also open questions.
\end{enumerate}

Another important aspect of compaction refers to the response to an
external perturbation, e.g. a change of the vibration intensity. It
happens that many of the peculiar dynamic behaviors exhibited by
granular materials under tapping are similar to those shown by
conventional structural glasses. For instance, both kind of systems
present slow relaxation, annealing properties, and hysteresis
effects. A necessary first step in order to understand these issues
is to study the response of the system to a small perturbation of
the conditions defining its steady state (linear response). Of
course, the obvious  external parameter to vary in the compaction
experiment is the vibration intensity characterizing the tapping
process, and the quantity whose response to analyze is the volume of
the system. In the process, all the system, with a constant number
of particles, is considered.

An additional reason to study linear response in granular systems,
is that fluctuation-dissipation relations obtained in molecular
systems in the context of linear response theory, are being used to
test the thermodynamic theory of granular matter
\cite{MyK02,CCyN06}. Nevertheless, given the non thermal character
of granular systems, even if the validity of the thermodynamic
theory as proposed by Edwards is assumed, there is no reason {\em a
priori} to expect the fluctuation-dissipation relations to apply.

The aim here is to investigate the above mentioned issues by means
of a very simple model for compaction \cite{ByP03a}. The model
dynamics is defined in such a way that it conserves the number of
particles in the system, i.e. it is a closed system. Its volume
changes due to the creation and destruction of empty sites or holes
in the lattice over which the system is defined. The model is
consistent with the Edwards thermodynamic theory, although of course
it is far away of any physically feasible system. Nevertheless, it
is interesting to qualitatively compare its predictions with the
experimental results and with the behavior obtained by means of
numerical simulation (Monte Carlo and molecular dynamics) of much
more sophisticated and realistic models.

The paper is organized as follows. In the next section the model to
be used in the following is formulated as a Markov process, and some
of its properties derived in ref. \cite{ByP03a} are shortly
summarized. In particular, the steady distribution characterizing
the long time limit of compaction is reminded and the Edwards
compactivity identified. From the steady distribution, explicit
expressions for some average properties are easily derived. Special
attention is given to the relationship between the second moment of
the volume fluctuations and the average volume, since it has been
used to compute the compactivity in experiments. The detailed shape
of the volume fluctuations is analyzed in Sec.\ \ref{s3}, where it
is predicted to be Gaussian in the limit of a large system, i.e.
with a large enough number of particles. This is verified by means
of Monte Carlo simulations of the model equations. The above model
predictions are compared with experimental findings and simulation
results for more realistic models in Sec.\ \ref{s4}. Special
emphasis is put on the discussion of the boundary conditions, e.g.
open versus closed systems, the local or global character of the
measurements, and the concept of effective temperature and its
relationship with the compactivity. In Sec. \ref{s5}, the linear
response of the model to a change in the tapping intensity is
investigated. The volume response function is computed by Monte
Carlo simulation, and a stretched exponential decay is found in the
relevant part of the relaxation. Finally, in the last section, a
short summary and some additional discussions are presented.

\section{The model and the steady distribution}
\label{s2}

Consider a one-dimensional lattice having $N+1$ particles and a
variable number of sites. The dynamics of the system is modeled
trying to capture, in a very simplified way, the relevant features
of compaction processes in tapped granular systems
\cite{KFLJyN95,NKPJyN97,NKBJyN98}. Then, a Markov process is
introduced with the elementary transitions given in Table
\ref{table1}, where the circles represent particles and the crosses
empty sites or holes, and $p$ and $q$ are arbitrary positive real
parameters \cite{ByP03a}. Time is measured in some arbitrary units
proportional to the number of taps. Only those particles involved
and/or conditioning each of the transitions are represented. It is
worth to point out that the equality of the rates for diffusion and
hole annihilation follows from the rational of the model
\cite{ByP03a}.

\begin{table}
\caption{\label{table1} Elementary transitions and their transition
rates for the lattice model. Particles and holes are represented by
circles and crosses, respectively.} \
\begin{ruledtabular}
\begin{tabular}{lccc}
Transition & Initial state & Final state & Rate\\
\hline
diffusion & OOXO & OXOO & $p/2$ \\
diffusion & OXOO & OOXO & $p/2$ \\
hole annihilation & OXOXO & OXOO & $p/2$ \\
hole annihilation & OXOXO & OOXO & $p/2$ \\
hole creation & OXOO & OXOXO & $q$ \\
hole creation & OOXO & OXOXO & $q$ \\
\end{tabular}
\end{ruledtabular}
\end{table}

The transitions in Table \ref{table1} define the modeled effective
dynamics connecting metastable configurations in a system being
tapped. These configurations are characterized by the property
that all the holes are isolated, i.e. surrounded by two particles
and, in order to enumerate them, a set of variables ${\bm n}
\equiv \{ n_{1},n_{2}, \ldots, n_{N} \}$ is introduced. By
definition, the variable $n_{i}$ takes the value $n_{i}=1$ if
there is a hole to its right, separating it from particle $i+1$.
On the other hand, $n_{i}=0$ if the site to the right of particle
$i$ is occupied by particle $i+1$.  It is assumed that there can
not be a hole either to the left of particle $1$ or to the right
of particle $N+1$. Note that this feature is preserved by the
dynamical events allowed in the model. Moreover, the dynamics does
not change the relative order of the particles in the lattice and,
therefore, the identity of the particles plays no role in
describing the configurations of the system. More details, as well
as a discussion of the physical motivation of the model, are given
in ref. \cite{ByP03a}.

Introduce $R_{i}{\bm n}$ as denoting the configuration obtained from
${\bm n}$ by modifying the value of $n_{i}$, keeping the same all
the other variables, i.e.,
\begin{equation}
\label{2.1} R_{i} {\bm n} \equiv \{ n_{1}, \ldots, n_{i-1},
1-n_{i},n_{i+1}, \ldots, n_{N} \} .
\end{equation}
Therefore, $R_{i}{\bm n}$ differs from ${\bm n}$ only in the status,
empty or occupied, of the site to the right of particle $i$. The
probability transition rates $W({\bm n}^{\prime}| {\bm n})$ from
configuration ${\bm n}$ to configuration ${\bm n}^{\prime}$ defining
the Markov process and represented in Table \ref{table1}, can be
written in an analytical form as:
\begin{enumerate}
\item {\em Diffusion}
\begin{equation}
\label{2.2} W(R_{i}R_{i+1} {\bm n} |{\bm n}) = \frac{p}{2} \left[
(1-n_{i})n_{i+1}+n_{i}(1-n_{i+1}) \right],
\end{equation}
\item {\em Hole annihilation}
\begin{equation}
\label{2.3} W(R_{i}{\bm n} | {\bm n}) = \frac{p}{2} \left(
n_{i-1}n_{i}+n_{i} n_{i+1} \right),
\end{equation}
\item {\em Hole creation}
\begin{equation}
\label{2.4} W(R_{i}{\bm n} | {\bm n}) =q (1-n_{i})(n_{i-1}+n_{i+1}).
\end{equation}
\end{enumerate}
This stochastic process can be proven to be irreducible. In
addition, its (unique) steady probability distribution $p^{(s)}({\bm
n})$ verifies the detailed balance condition and is given by
\cite{ByP03a}
\begin{equation}
\label{2.5} p^{(s)}({\bm n}) = \frac{\gamma^{-N_{H}}}{Z(\gamma)}\, ,
\end{equation}
where $\gamma = p/ 2 q$ and
\begin{equation}
\label{2.6} N_{H}({\bm n}) = \sum_{i=1}^{N} n_{i}
\end{equation}
is the total number of holes of the lattice configuration ${\bm n}$.
Finally, $Z(\gamma)$ is the `partition function' following from the
normalization condition,
\begin{equation}
\label{2.7} Z(\gamma) = \sum_{\bm n} \gamma^{-N_{H}} =
\sum_{N_{H}=1}^{N} \Omega(N_{H}) \gamma^{-N_{H}},
\end{equation}
with $\Omega(N_{H})$ being the number of metastable configurations
having exactly $N_{H}$ holes. A standard combinatorial reasoning
yields
\begin{equation}
\label{2.8} \Omega (N_{H})= \frac{N!}{N_{H}! (N-N_{H})!}.
\end{equation}

From Eqs.\ (\ref{2.7}) and (\ref{2.8}) it follows that
\begin{equation}
\label{2.9} Z(\gamma)= \left( 1 +\frac{1}{\gamma} \right)^{N} -1,
\end{equation}
that for large $N$ implies
\begin{equation}
\label{2.10} \ln Z(\gamma) \approx N \ln \left( 1 +\frac{1}{\gamma}
\right).
\end{equation}
To be more precise, the required limit leading from Eq.\ (\ref{2.9})
to Eq.\ (\ref{2.10}) is $N \gg 1$ and $N \gg \gamma$, since if
$\gamma$ is of the same order as $N \gg 1$, the correct limiting
form for the partition function is
\begin{equation}
\label{2.11} Z (\gamma) \approx e^{N/\gamma}-1.
\end{equation}
In the following, attention will be restricted to systems having a
number of particles such that Eq.\ (\ref{2.10}) applies.

Once $\ln Z(\gamma)$ is known in a analytical form, it is a simple
task to compute average values characterizing the macroscopic state
of the system. Also, the probability distribution for the number of
holes in the steady state, $P^{(s)}(N_{H})$, is easily derived,
\begin{equation}
\label{2.12} P^{(s)}(N_{H}) = \sum_{\bm n} p^{(s)} ({\bm n})
\delta_{N_{H},\sum n_{i}} = \frac{\Omega (N_{H})
\gamma^{-N_{H}}}{Z(\gamma)}.
\end{equation}
In the second equality above, $\delta_{n,m}$ is the Kronecker delta.
The associated moment generating function \cite{vK92} is
\begin{eqnarray}
\label{2.13} G_{N_{H}} (k) & \equiv & \sum_{N_{H}=1}^{N} P^{(s)}
(N_{H}) e^{ikN_{H}} = \frac{1}{Z(\gamma)} \sum_{N_{H}=1}^{N}
\Omega (N_{H}) \left( \gamma e^{-ik} \right)^{-N_{H}} \nonumber \\
&=& \frac{Z(\gamma e^{-ik})}{Z(\gamma)}.
\end{eqnarray}
Thus the cumulant generating function \cite{vK92}  of the variable
$N_{H}$ is
\begin{equation}
\label{2.14} \ln G_{N_{H}}(k) \approx N \ln
\frac{\gamma+e^{ik}}{\gamma+1},
\end{equation}
where use has been made of Eq.\ (\ref{2.10}). From this expression,
the steady average of the number of holes $\langle N_{H}
\rangle^{(s)}$ and its variance $\sigma_{N_{H}}^{2}$ are directly
computed,
\begin{equation}
\label{2.15}  \langle N_{H}\rangle^{(s)} = \frac{1}{i} \left[
\frac{\partial \ln G_{N_{H}} (k)}{\partial k} \right]_{k=0} =
\frac{N}{\gamma+1},
\end{equation}
\begin{equation}
\label{2.16} \sigma_{N_{H}}^{2} \equiv \langle N_{H}^{2}
\rangle^{(s)} - \langle N_{H} \rangle ^{(s)2} = - \left[
\frac{\partial ^{2} \ln G_{N_{H}}(k)}{\partial k^{2}} \right]_{k=0}
= \frac{N \gamma}{(\gamma+1)^{2}}.
\end{equation}
From Eq.\ (\ref{2.15}), it is seen that the steady state gets more
compact as $\gamma$ increases. On the other hand, it is found that
the relaxation of the system becomes slower \cite{ByP03a}. This is
consistent with the physical interpretation of the model, where
$\gamma^{-1}$ is understood to play a role analogous to the tapping
intensity  in granular experiments \cite{KFLJyN95,NKBJyN98}. Now, in
order to get a canonical distribution of the form proposed by
Edwards and coworkers \cite{EyO89}, the `compactivity' $X$ is
defined by
\begin{equation}
\label{2.17} X \equiv \frac{1}{\ln \gamma},
\end{equation}
so the steady distribution in Eq.\ (\ref{2.5}) can be rewritten as
\begin{equation}
\label{2.18} p^{(s)}({\bm n})=\frac{e^{-N_{H}/X}}{Z(X)}, \quad
Z(X)=\left( 1 +e^{-1/X} \right)^{N}.
\end{equation}
In terms of the compactivity, Eqs. (\ref{2.15}) and (\ref{2.16})
take the form
\begin{equation}
\label{2.19} \langle N_{H}\rangle^{(s)} =\frac{N}{e^{1/X}+1},
\quad \sigma_{N_{H}}^{2} =\frac{Ne^{1/X}}{\left( e^{1/X}+1
\right)^{2}},
\end{equation}
respectively. Combination of the above two equations gives
\begin{equation}
\label{2.19a} \frac{\sigma_{N_{H}}}{ \langle N_{H} \rangle^{(s)}}
=\frac{e^{1/2 X}}{\sqrt{N}} = \sqrt{\frac{\gamma}{N}},
\end{equation}
showing the typical $N^{-1/2}$ dependence of the relative
fluctuations of an extensive quantity on the number of particles of
the system.

Since $p$ and $q$ can take arbitrary positive real values, it is
clear from the definition of $\gamma$ given below Eq.\ (\ref{2.5}),
that its range of variation is $0 <\gamma < \infty$. Consequently,
$X$ can take any real value, i.e. $- \infty < X < + \infty$. More
precisely, when $\gamma$ is increased monotonically from $0$ to
$\infty$, the compactivity $X$ decreases from $0^{-}$ to $-\infty$
(for $\gamma= 1^{-}$), then jumping to $+ \infty$ ($\gamma=1^{+}$)
and decreasing again towards $0^{+}$. Then, there is a parameter
region, namely $p<2q$, or $0<\gamma <1$, in which negative values of
the compactivity occur. With reference to the dynamics of the model,
this happens when the transition rate for the hole creation
processes is larger than that for hole annihilation processes.

Equation (\ref{2.18}) lead to the relation
\begin{equation}
\label{2.20} \frac{\partial}{\partial (1/X)} \langle N_{H}
\rangle^{(s)} =- \sigma_{N_{H}}^{2},
\end{equation}
and considering a series of experiments with the same number of
particles in which the average number of holes is changed between
$\langle N_{H} \rangle_{1}^{(s)}$ and $\langle N_{H}
\rangle^{(s)}_{2}$,
\begin{equation}
\label{2.21} - \int_{\langle N_{H} \rangle_{1}^{(s)}}^{\langle N_{H}
\rangle_{2}^{(s)}} d \langle N_{H}\rangle^{(s)}\,
\frac{1}{\sigma_{N_{H}}^{(s)2}} = \frac{1}{X_{2}}-\frac{1}{X_{1}}.
\end{equation}
Here $X_{1}$ and $X_{2}$ are the initial and final values of the
compactivity, respectively. The generalization of this equation for
three-dimensional systems in the context of the Edwards theory was
first noted by Nowak et al. \cite{NKBJyN98}, and used in ref.
\cite{SGyS05} to obtain the compactivity as a function  of the
volume fraction from experimental data of a system submitted to
pulses. More will be said about this in Sec.\ \ref{s4}.

\section{Volume fluctuations}
\label{s3} The length $L$ of the system, measured in units of the
distance between consecutive sites, is
\begin{equation}
\label{3.1} L = N +N_{H}.
\end{equation}
In the steady state, the average value of $L$ is
\begin{equation}
\label{3.2} \langle L \rangle^{(s)} =N + \langle N_{H} \rangle^{(s)}
= \frac{N(\gamma+2)}{\gamma+1}.
\end{equation}
It is also convenient to introduce the length per particle $l \equiv
L/N$, so $\langle l \rangle^{(s)} = (\gamma+2)/(\gamma+1)$, or
\begin{equation}
\label{3.3} \gamma=\frac{2-\langle l \rangle ^{(s)}}{ \langle l
\rangle^{(s)}-1}.
\end{equation}
This can be transformed into a relationship between $X$ and $\langle
l \rangle^{(s)}$ by means of Eq. (\ref{2.17}),
\begin{equation}
\label{3.4} \frac{1}{X}= \ln \frac{2 - \langle l
\rangle^{(s)}}{\langle l \rangle ^{(s)}-1}.
\end{equation}
It is seen that $1/X$ is a monotonically decreasing function of $
\langle l \rangle^{(s)}$, vanishing for $\langle l
\rangle^{(s)}=3/2$ that, of course, corresponds to $\gamma=1$.
Therefore, it is concluded that $X>0$ for $\langle l \rangle
^{(s)}<3/2$, while $X<0$ for $ \langle l \rangle ^{(s)}> 3/2$.
Negative compactivities correspond to the loosest configurations and
have also been obtained in other models of compaction \cite{MyP97}.
The inverse of the compactivity does not vanish in the limit of the
loosest possible configuration, but it tends to $- \infty$.
Obviously, the existence of negative values of the compactivity does
not violate any general physical principle and does not imply any
instability of the steady state. They are a consequence of the fact
that the number of configurations $\Omega(N_{H})$ is not a monotonic
increasing function of the length of the system (or $N_{H}$), but it
exhibits a maximum, going afterwards to unity for the maximum value
of the number of holes, $N_{H}=N$. Equivalently, the possibility of
steady states with $X<0$ is directly related with the length $L$ of
the system having a finite upper bound. From a formal point of view,
the situation is similar to the existence of negative temperatures
in systems of nuclear spins in certain paramagnetic crystals placed
in an external magnetic field.

From Eq. (\ref{2.16}), the variance of the total length can be
written as
\begin{equation}
\label{3.5} \sigma_{L}^{2} = \sigma_{N_{H}}^{2}= N \left( \langle  l
\rangle^{(s)}-1 \right) \left( 2 -\langle l \rangle^{(s)} \right).
\end{equation}
In order to get a more detailed information of the length
fluctuations, introduce the stochastic variable $Y$ by
\begin{equation}
\label{3.6} Y = \frac{L - \langle L \rangle^{(s)}}{\sigma_{L}} =
\frac{N_{H}-\langle N_{H} \rangle^{(s)}}{\sigma_{N_{H}}}.
\end{equation}
The moment generating function of $Y$ is
\begin{eqnarray}
\label{3.7} G_{Y}(k) & \equiv & \langle e^{ikY} \rangle^{(s)} =
e^{-ik \frac{\langle N_{H} \rangle^{(s)}}{\sigma_{N_{H}}}} \langle
e^{ik \frac{ N_{H}}{\sigma_{N_{H}}}} \rangle^{(s)} \nonumber \\
&=& e^{-ik \frac{\langle N_{H} \rangle^{(s)}}{\sigma_{N_{H}}}}
\left[ G_{N_{H}}(q) \right]_{q= k/\sigma_{N_{H}}},
\end{eqnarray}
where $G_{N_{H}}$ was defined in Eq. (\ref{2.13}). Then, using Eqs.\
(\ref{2.14})-(\ref{2.16}),
\begin{equation}
\label{3.8} \ln G_{Y}(k)= -ik N^{1/2} \gamma^{-1/2}+N \ln \left\{ 1
+ \gamma^{-1}  \exp \left[i N^{-1/2} \gamma^{-1/2}(1+\gamma)k
\right] \right\} - N \ln \frac{1+\gamma}{\gamma},
\end{equation}
that in the limit $ N \gg 1$ reduces to
\begin{equation}
\label{3.9} \ln G_{Y}(k) \sim - \frac{k^{2}}{2}\, .
\end{equation}
This proves that, for a large enough number of particles, the
stochastic variable $Y$ obeys a Gaussian distribution with vanishing
average value and unit variance. Formally, this is equivalent to
saying that the probability density $\omega_{l}(l)$ for the variable
$l$ is also Gaussian with average $\langle l \rangle^{(s)}$ and
standard deviation $\sigma_{l}=\sigma_{L}/N$.

\begin{figure}
\includegraphics[scale=0.45]{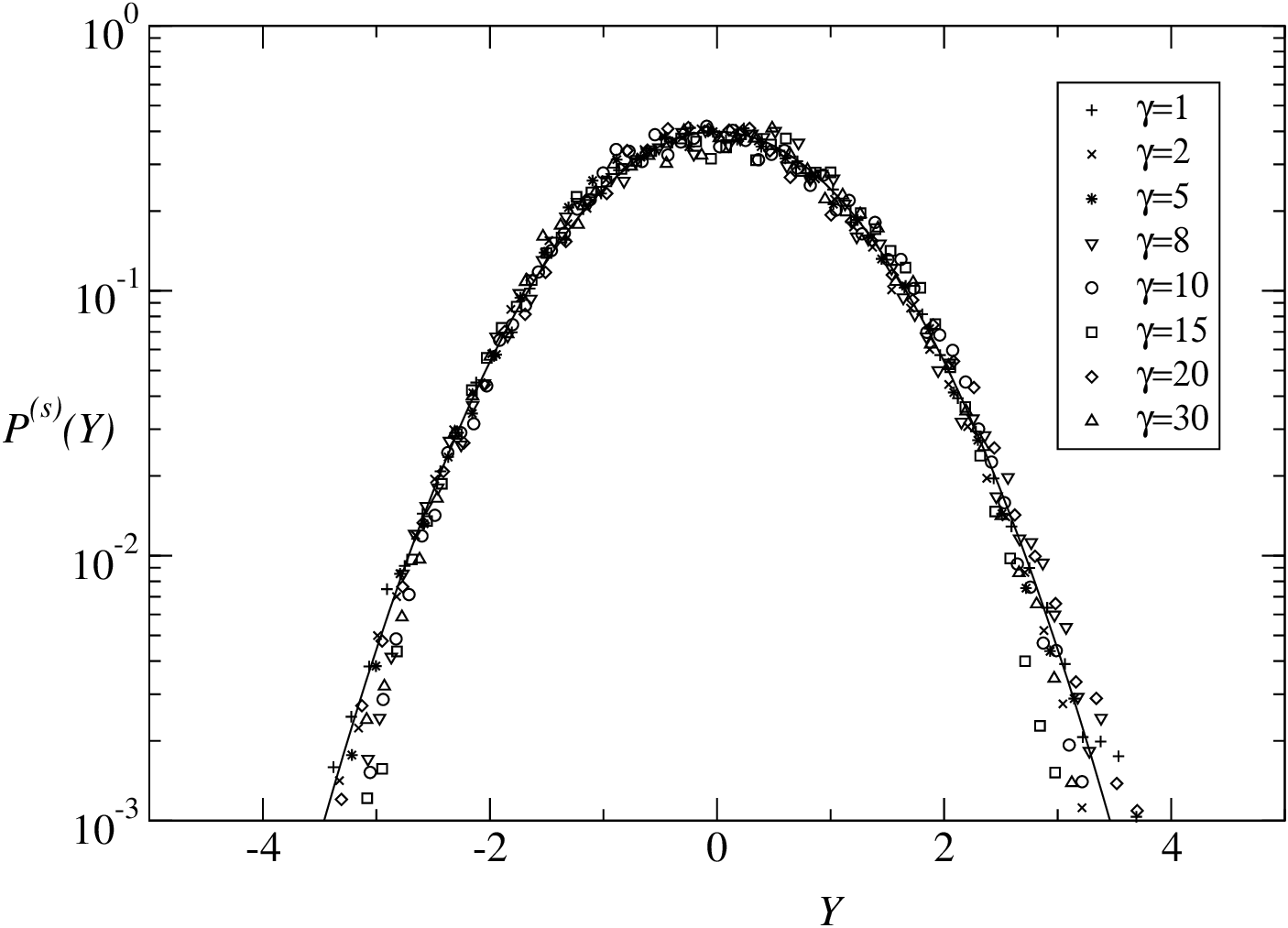}
\caption{Steady probability distribution $P^{(s)}(Y)$ of the
dimensionless variable $Y$ defined in Eq.\ (\protect{\ref{3.6}}).
The symbols are Monte Carlo simulation results obtained with
different values of $\gamma$, as indicated in the insert. The
solid line is the Gaussian theoretical prediction. \label{fig1}}
\end{figure}

A numerical check of the Gaussian character of $Y$ for large values
of $N$ is presented in Fig.\ \ref{fig1}, where Monte Carlo
simulation results for the steady probability distribution
$P^{(s)}(Y)$  are shown for several values of $\gamma$. The number
of particles used is $N=10^{5}$. The solid line is the Gaussian
distribution of zero average and unit variance. An excellent
agreement is observed, even for rather small values of $P^{(s)}(Y)$.

In order to compare the results derived for the model in this
Section with some experimental findings, it is convenient to
consider also the volume fraction $\phi$ defined as $\phi \equiv
l^{-1}$. As a consequence of $\sigma_{l}$ being proportional to
$N^{-1/2}$, the probability density for $\phi$,
$\omega_{\phi}(\phi)$ is also Gaussian, again in the limit of
large $N$, with average and variance given by
\begin{equation}
\label{3.10} \langle \phi \rangle^{(s)}= \langle l \rangle^{(s)-1},
\end{equation}
and
\begin{equation}
\label{3.11} \sigma_{\phi}= \frac{\sigma_{l}}{\langle l
\rangle^{(s)2}} = \frac{\left[ \left( \langle l  \rangle ^{(s)} - 1
\right) \left( 2 - \langle l \rangle^{(s)} \right)
\right]^{1/2}}{N^{1/2} \langle l \rangle^{(s)2}}\, ,
\end{equation}
respectively. Thus Eq.\  (\ref{3.4}) is, for large $N$, equivalent
to
\begin{equation}
\label{3.11} \frac{1}{X}= \ln \frac{2 \langle \phi
\rangle^{(s)}-1}{1-\langle \phi \rangle^{(s)}}\, .
\end{equation}

\section{Comparison of the model predictions with some
experimental results} \label{s4} Of course, given the simplicity
of the model discussed here and its one-dimensional character, it
can not be expected to quantitatively reproduce the experimental
and simulation results. Nevertheless, it is interesting and
illuminating to qualitatively analyze those results in the light
of the model predictions.

The first experimental investigation of the volume fluctuations in
the steady state of a tapped granular medium seems to be the one
reported in ref. \cite{NKBJyN98}. More exactly, the property
considered in that work is the density in a given region, located at
a certain height of the system. Then, the volume is kept constant
and the measured density changes are due to variations in the number
of particles within the volume considered. The results indicated
that the probability distribution of the density fluctuations
exhibited in the majority of the cases a Gaussian shape, as  found
in the model, although some significant non-Gaussian deviations were
also found, particularly near the bottom of the vibrated column.
Moreover, the dependence on the number of particles of the relative
variance was also the same as in the model, i.e $N^{-1/2}$ as shown
in Eq.\ (\ref{2.19a}).

Variations in the inverse of the compactivity were also measured, by
using the relationship between the second moment of the volume
fluctuations $\sigma_{V}^{2}$ and the compactivity following from
Edwards' theory, i.e. the generalization of Eq.\ (\ref{2.21}). To do
so, the authors considered the specific volume $v$ defined as the
inverse of the packing fraction. This corresponds to the quantity
$l$ in our model. The reported results are
\begin{equation}
\label{4.1} \sigma_{v}^{2} = a +b \langle v \rangle^{s},
\end{equation}
\begin{equation}
\label{4.2} \frac{1}{X} \propto \ln (a+b \langle v \rangle^{(s)}) +
\text{constant},
\end{equation}
where $a$ and $b$ are two dimensionless parameters. Taking into
account that the available range of experimental data is rather
limited and located in the proximity of the close packing limit, it
can be fairly concluded that the above expressions are in
qualitative agreement with the predictions of the model, namely the
expressions for $\sigma_{l}$ and $X^{-1}$, Eqs. (\ref{3.11}) and
(\ref{3.4}) respectively. Nevertheless, to put this comparison in a
fair proper context, the following two comments seem appropriated:
\begin{enumerate}

\item In ref. \cite{NKBJyN98}, the Edwards theory is considered in
a {\em local} form, i.e. it is assumed that the state of each
subregion of the system in which measurements are carried out is
described by a canonical distribution with a different value of the
compactivity. Although this might be a sensible idea, it is clearly
an extension of the theory as originally formulated by Edwards.

\item Moreover, what is measured in the reported experiments are
the fluctuations in the number of particles at constant volume. This
is different from the fluctuations in the volume at constant number
of particles, although both can be associated to density
fluctuations. In fact, it can be shown that the quantity actually
obtained from the second moment of the number of particles
fluctuations by means of a relationship similar to Eq.\ (\ref{2.21})
is not the compactivity, but some kind of `fugacity' \cite{ByP03a}.
Although both quantities are simply related and perhaps their values
are very close in real granular systems, they are conceptually quite
different.
\end{enumerate}

Another experimental study of the steady volume fluctuations in a
granular medium has been carried out by Schr\"{o}ter {\em et al.}
\cite{SGyS05}. This work might look as quite similar to the one
discussed above, but they differ in important aspects, both in the
methodology and in the results. In ref. \cite{SGyS05}, the
fluctuations of the {\em total} volume of the system were measured,
although the results are expressed in terms of the volume fraction
$\phi$.  A Gaussian distribution was found in all the reported
cases. Also, the ratio between the standard deviation and the
average turned out to be proportional to the inverse of the square
root of the number of particles. Both results are the same as
derived above for the model considered here. Nevertheless, the
variance of the volume fraction fluctuations  did not show the
analogous of the simple linear behavior reported in ref.\
\cite{NKBJyN98} that, as it has been already mentioned, is
consistent with the linear behavior predicted by Eq.\ (\ref{3.11})
near the close packing limit. Instead, it presents a well defined
minimum. A first, naive explanation is provided by the fact that a
wider density range is analyzed in \cite{SGyS05} than in
\cite{NKBJyN98}. Then, it could be expected that the linear law in
the latter corresponds to an approximate description of a small
window of the density interval considered in the former, namely to
one with the largest volume fraction values. Even assuming this
explanation, the origin of the minimum would remain as an open
question.

The result obtained with the model, Eq.\ (\ref{3.11}), when
considered over the density range $  1/2 \leq \langle \phi
\rangle^{(s)} \leq 1$, also shows a non-monotonic behavior, but
presenting a maximum instead of a minimum exhibited by the
experimental data in ref. \cite{SGyS05}. Then, it seems plausible to
conclude that the existence of the minimum, if confirmed, must be
due to some effects that are not captured by the simple model
considered here. Given that it appears for high values of the
density, it is tempting to speculate whether the increase of the
volume fluctuations in that region, to the right of the minimum, is
an indication of the presence of some `singular' behavior
associated, for instance, with the existence of a maximum random
packing, a phenomenon that is not present in the model.

The Edwards compactivity $X$ was also determined by Schr\"{o}ter
{\em et al.} by means of the granular version of the equilibrium
fluctuation-dissipation theorem. To get explicit values for $X$,
these authors assume that the inverse of the compactivity vanishes
in the random loose packing limit. Since $X$  is a monotonic
decreasing function of the volume fraction, this boundary condition
implies that it is always positive. As it has already been
indicated, there is nothing in the granular statistical mechanics
theory requiring this to be true. If attention is restricted to the
parameter region in which the model has positive compactivity, i.e.
$\langle \phi \rangle^{(s)} >2/3$, the shape of the curve $X(\langle
\phi \rangle^{(s)})$, Eq. (\ref{3.11}), is similar to the one
reported in Fig. 4(c) of ref. \cite{SGyS05}.

Although the recent results  presented in ref. \cite{CCyN06} have
not been obtained experimentally, but by means of molecular dynamics
and Monte Carlo simulations, they deserve special attention because
of their relevance and the interest of the analysis carried out. In
particular, interest is focused on a parameter defined as the
'configurational granular temperature', instead of the compactivity,
although some close relationship between both parameters is claimed
to hold.  A model system of grains with normal and tangential forces
is considered. The instantaneous volume fraction is measured in the
steady state and analyzed as a function of its average value. To
measure them, a region in the bulk of the system is considered. It
is important to stress that this region is defined in such a way
that both its volume and its number of particles  fluctuate in time.
Consistently with all the results discussed above and the model
predictions, Gaussian fluctuations of the volume fraction are also
found in the simulations.

The analysis of the results in \cite{CCyN06} differs, at least
conceptually, of the thermodynamic theory as proposed initially by
Edwards \cite{EyO89}. Instead of considering that the macroscopic
parameter characterizing the state of the system is the volume,
the energy is used like in normal molecular systems. More
precisely, the probability $p^{(s)}_{r}$ of finding the system in
the blocked metastable state $r$ of energy $E_{r}$ is assumed to
be given by \cite{CyN01,FNyC02}
\begin{equation}
\label{4.3} p^{(s)}_{r} \propto e^{-E_{r}/T_{\text{conf}}},
\end{equation}
where $T_{\text{conf}}$ is referred to as the configurational
granular temperature of the system. The above expression is to be
compared with Edwards' proposal, see also Eq.\ (\ref{2.18}),
\begin{equation}
\label{4.4} p_{r}^{(s)} \propto e^{-V_{r}/X},
\end{equation}
$V_{r}$ being the volume of the system in the configuration $r$.
Both expressions only become equivalent if, for the relevant
configurations, i.e. those with a non-vanishing probability, the
energy and the volume are proportional \cite{EPAPS}, with a
configuration-independent constant. Consequently, the
configurational temperature $T_{\text{conf}}$ and the compactivity
$X$ also would be proportional in this case.

In the paper, the authors plot  (fig. 4) the standard deviation
$\sigma_{\phi}$ of the volume fraction fluctuations as a function of
the steady average volume fraction $\langle \phi \rangle^{(s)}$. In
the plotted interval, a linear behavior of  $\sigma_{\phi} (\langle
\phi \rangle^{(s)})$ is clearly identified. Then, by using the
relationship between $\sigma_{\phi}$ and the compactivity $X$, an
analytical expression of the form \cite{ByP07b}
\begin{equation}
\label{4.5} \frac{1}{X}= c+f(\langle \phi \rangle^{(s)})
\end{equation}
is obtained. Here, the explicit form of the function $f(\langle \phi
\rangle^{(s)})$ is known, while $c$ is an unknown constant. If
$T_{\text{conf}} = d  X$, with $d$ a constant, Eq. (\ref{4.5})
yields
\begin{equation}
\label{4.6} f(\langle \phi \rangle^{(s)})=\frac{d}{T_{\text{conf}}}-
c.
\end{equation}

In ref. \cite{CCyN06}, a plot of the volume fraction as a function
of the configurational temperature, obtained by means of Monte Carlo
simulation, is also provided (fig. 5). Consistently, this plot
covers the same density range as the one giving the fluctuations.
Therefore, it is posible to check whether the linearity between
$f(\langle \phi \rangle^{(s)})$ and $T_{\text{conf}}^{-1}$ predicted
by Eq.\ (\ref{4.6}) is verified. When this is done, a systematic but
quite small deviation from linearity is observed. Then, taking into
account possible numerical errors as well as the fact that the
measurements are carried out in such a way that both the volume
considered and the number of particles inside it fluctuate, no
definite conclusion can be reached on the equivalence of the
canonical distribution in energy, Eq. (\ref{4.3}), and volume, Eq.
(\ref{4.4}) \cite{ByP07b}. This is an important point, both from
fundamental and applied perspectives and deserves further attention.

\section{Linear response. Volume autocorrelation function} \label{s5}
In this Section, the linear response of the model to a perturbation
of the parameter $\gamma$, introduced below  Eq.\ (\ref{2.5}) will
be studied. This parameter measures the ratio of the transition
rates of the processes associted with hole annihilation and
creation, respectively. Because of the definition in Eq.\
(\ref{2.17}), a change in $\gamma$ is equivalent to modifying the
compactivity $X$. In the context of real experiments, it might
correspond to varying the vibration intensity, defined as the ratio
between the peak acceleration of the shakes and the gravity
acceleration \cite{KFLJyN95}.

The typical relaxation experiment to be considered is as follows.
The system is initially prepared in a uniform density configuration
corresponding to a low density state, namely with $N_{H}=N$. Then,
it is allowed to evolve with a value $\gamma + \Delta \gamma$ of the
rate ratio, until it reaches a steady state with a probability
distribution $p^{(s)} ({\bm n};\gamma+ \Delta \gamma)$. At a given
moment, taken as the time origin $t=0$, the rate ratio is
instantaneously changed to the value $\gamma$. It is assumed that
$|\Delta \gamma| \ll \gamma$. Then, the relaxation of the system to
a new steady state $p^{(s)}({\bm n};\gamma)$ is followed.

The probability distribution function of the system at $t=0$ is
\begin{equation}
\label{5.1} p({\bm n},t=0)= p^{(s)}({\bm n}; \gamma+\Delta
\gamma)\approx p^{(s)} ({\bm n}; \gamma) +\frac{\partial
p^{(s)}({\bm n};\gamma)}{\partial \gamma} \Delta \gamma.
\end{equation}
Using Eq.\ (\ref{2.18}), it is obtained
\begin{equation}
\label{5.2} \frac{\partial p^{(s)}({\bm n};\gamma)}{\partial
\gamma} = \frac{\partial p^{(s)}({\bm n};\gamma)}{\partial X}
\frac{ dX}{d \gamma} = \frac{N_{H}-\langle N_{H};\gamma
\rangle^{(s)}}{X^{2}}\, p^{(s)}({\bm n}; \gamma) \frac{dX}{d
\gamma}\, .
\end{equation}

The average value of a function $F({\bm n})$ of the system
configuration at time $t>0$ is, by definition,
\begin{equation}
\label{5.3} \langle F,t \rangle \equiv  \sum_{\bm n} F({\bm n})
p({\bm n},t);
\end{equation}
where the probability distribution $p({\bm n},t)$ is obtained from
$p({\bm n},t=0)$ through
\begin{equation}
\label{5.4} p({\bm n},t)= \sum_{{\bm n}^{\prime}} p({\bm n},t | {\bm
n}^{\prime},0) p({\bm n}^{\prime},t=0).
\end{equation}
Here,  $p({\bm n},t | {\bm n}^{\prime},0)$ is the conditional
probability that the configuration of the system is ${\bm n}$ at
time $t$ given it was ${\bm n}^{\prime}$ at $t=0$. By construction,
$p^{(s)}({\bm n};\gamma)$ is stationary for the dynamics occurring
for $t>0$, i.e. with rate ratio $\gamma$,
\begin{equation}
\label{5.5} \sum_{{\bm n}^{\prime}}  p({\bm n},t | {\bm
n}^{\prime},0) p^{(s)}({\bm n}^{\prime}; \gamma) = p^{(s)}({\bm
n};\gamma).
\end{equation}
Taking this into account and using Eqs.\ (\ref{5.1}) and
(\ref{5.2}), Eq.\ (\ref{5.3}) becomes
\begin{equation}
\label{5.6} \langle F,t \rangle =  \langle F;\gamma \rangle^{(s)}
+\frac{1}{X^{2}} \frac{dX}{d \gamma}\, \langle \Delta F(t) \Delta
L(0);\gamma \rangle^{(s)} \Delta \gamma,
\end{equation}
where $ \langle \Delta F(t) \Delta L(0);\gamma \rangle^{(s)}$ is the
steady time correlation function of $F({\bm n})$ and the volume $L$
defined as
\begin{equation}
\label{5.7} \langle \Delta F(t) \Delta L(0);\gamma \rangle^{(s)} =
\sum_{\bm n} \sum_{{\bm n}^{\prime}} \Delta F({\bm n}) p({\bm n},t |
{\bm n}^{\prime},0) \Delta L({\bm n}^{\prime}) p^{(s)} ({\bm
n}^{\prime};\gamma),
\end{equation}
with
\begin{equation}
\label{5.8} \Delta F({\bm n}) =F({\bm n})-\langle F;\gamma
\rangle^{(s)},
\end{equation}
\begin{equation}
\label{5.9} \Delta L({\bm n}) =L({\bm n})-\langle L;\gamma
\rangle^{(s)}.
\end{equation}

\begin{figure}
\includegraphics[scale=0.45]{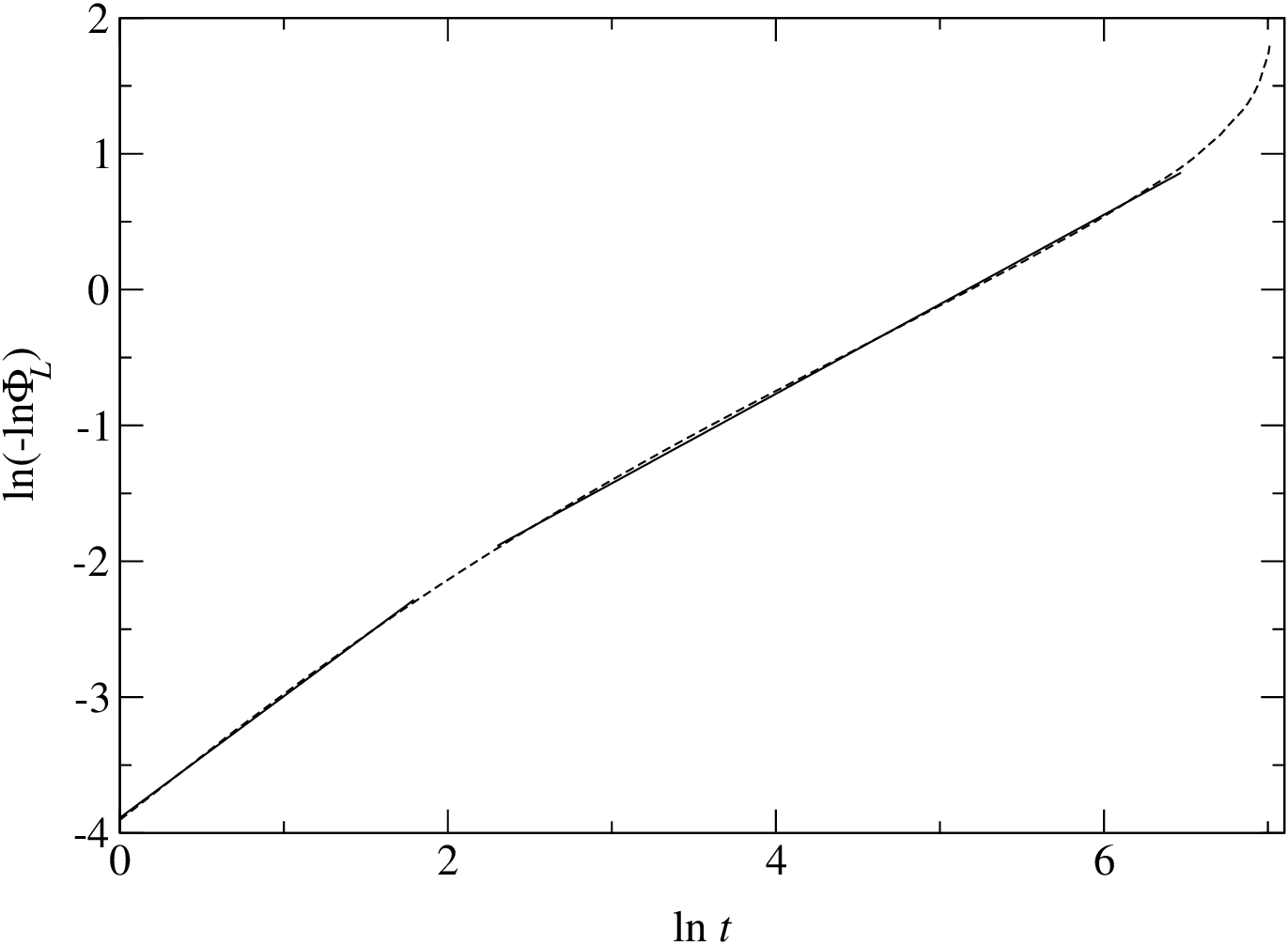}
\caption{Double logarithm plot of the normalized response function
for the volume as a function of the logarithm of the time for
$\gamma=5$. The latter is measured in dimensionless units
proportional to the number of Taps. The dashed line is the Monte
Carlo simulation result, while the two solid lines are linear fits
as discussed in the text.\label{fig2}}
\end{figure}

\begin{figure}
\includegraphics[scale=0.45]{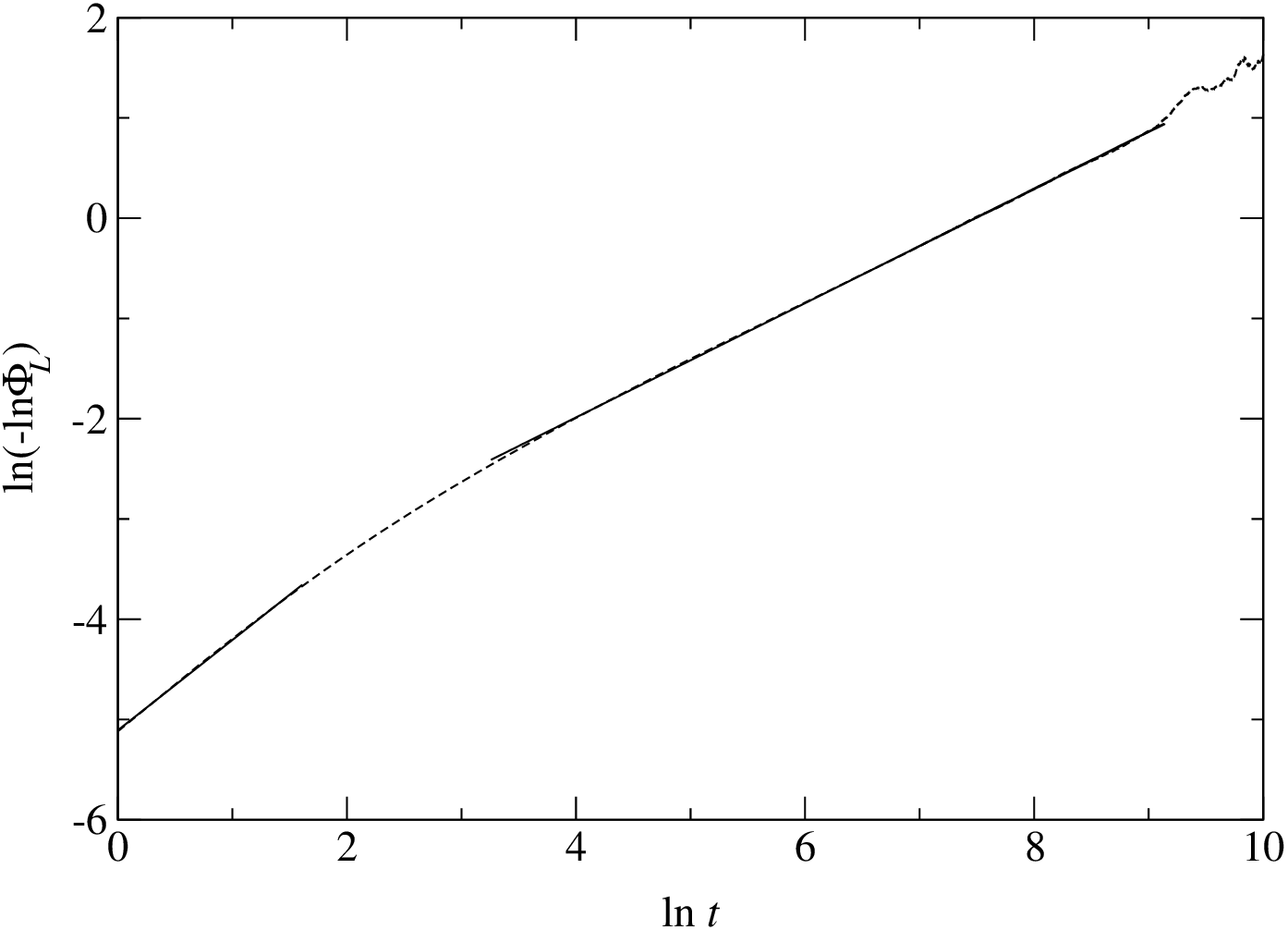}
\caption{The same as Fig. \protect{\ref{fig2}}, but for $\gamma=15$.
\label{fig3}}
\end{figure}

\begin{figure}
\includegraphics[scale=0.45,clip= ]{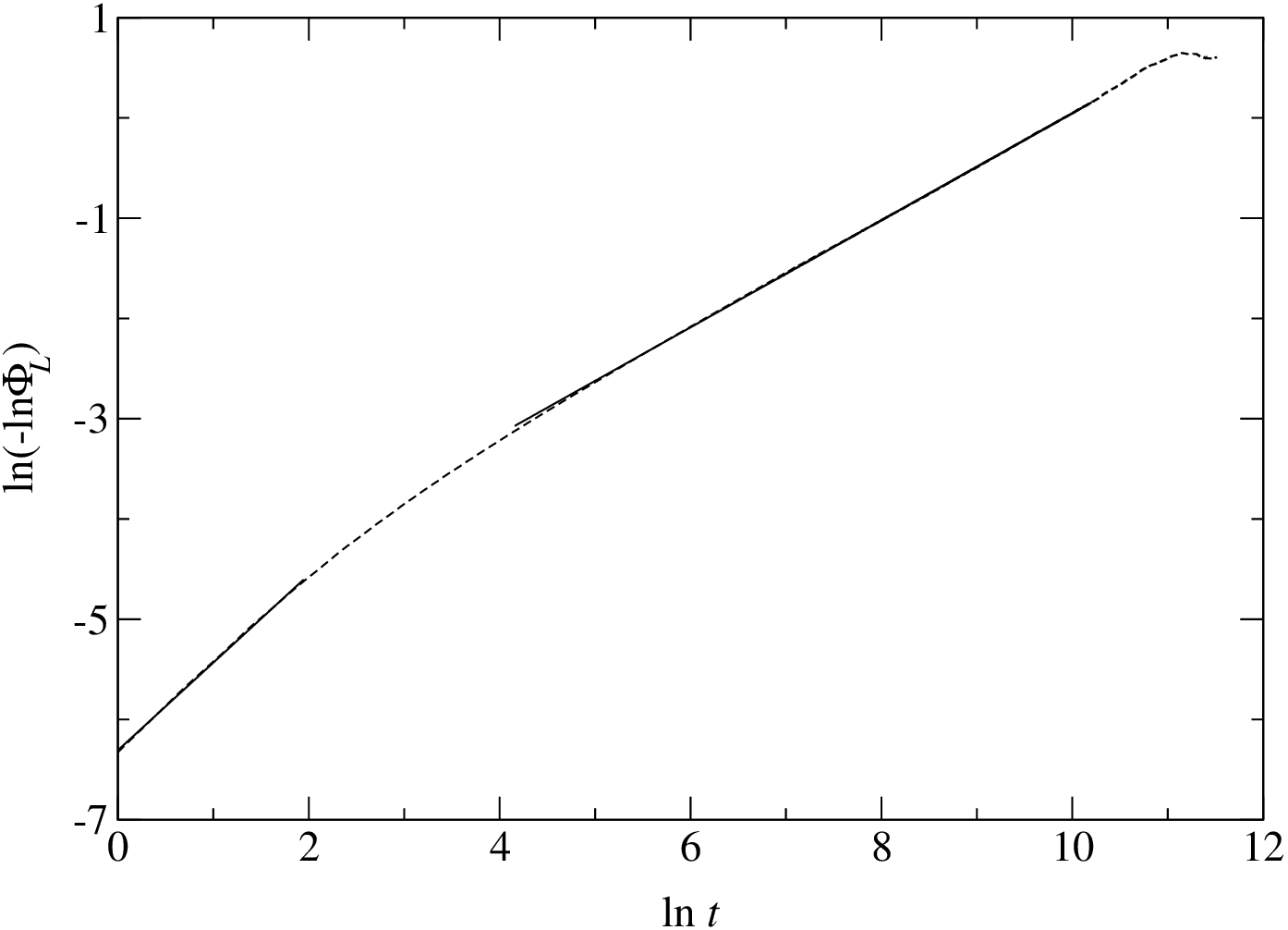}
\caption{The same as Fig. \protect{\ref{fig2}}, but for $\gamma=50$.
\label{fig4}}
\end{figure}

The normalized relaxation function for the property $F$,
$\Phi_{F}(t)$ is
\begin{equation}
\label{5.10} \Phi_{F}(t) \equiv \frac{\langle F;t \rangle- \langle
F; \gamma \rangle^{(s)}}{\langle F,0\rangle-\langle F;\gamma
\rangle^{s}}= \frac{\langle \Delta F(t) \Delta L(0);\gamma
\rangle^{(s)}}{\langle \Delta F \Delta L; \gamma\rangle^{(s)}}\, .
\end{equation}
For the special choice $F=L$,
\begin{equation}
\label{5.11} \Phi_{L}(t) =\frac{\langle \Delta L(t) \Delta
L(0);\gamma \rangle^{(s)}}{\sigma_{L}^{2}}.
\end{equation}
This expression is an exact consequence of modelling the dynamics of
the system by means of a Markov process and of the canonical form of
the steady probability distribution. The Monte Carlo simulation
results for the relaxation function  to be presented in the
following, have been obtained by using Eq.\ (\ref{5.11}). Figures
\ref{fig2}, \ref{fig3}, and \ref{fig4} show $\Phi_{L}(t)$ for three
values of the rate ratio, namely $\gamma=5$, $15$, and $50$,
respectively. Other values have also been investigated, and the
results are consistent with the comments carried out below. In the
figures the quantity $\ln (-\ln \Phi_{L})$ has been plotted as a
function of $\ln t$. It turns out that this is a convenient
representation to identify the shape of the decay of $\Phi_{L}(t)$.
Note that in this representation, a straight line corresponds to a
stretched exponential or Kohlrausch-Willians-Watts (KWW) function,
\begin{equation}
\label{5.12} \Phi_{KWW} (t) = \exp \left[ -\left( \frac{t}{\tau}
\right)^{\beta} \right],
\end{equation}
where $\tau$ and $\beta$  are the two parameters identifying the
function.

In the three figures, two different regions are easily identified.
For `short' times, there is a linear region with slope $\beta
\approx 1$. i.e. an exponential decay. Afterwards, there is a region
of `intermediate' times in which another linear behavior shows up,
but now with $\beta <1$. The initial region extends over a time
interval, $\ln t \alt 2$, that is roughly independent of $\gamma$.
In this time interval, the response function $\Phi_{L}$ decays a
small amount. Moreover, this initial decay becomes smaller and
smaller as the value of $\gamma$ increases. On the other hand, in
the second time window in which the response function shows a KWW
behavior, most of the relaxation occurs. For $\gamma = 5$, Fig.\
(\ref{fig2}), the fit to a KWW function with $\beta \simeq 0.659$ is
quite good for the time interval $3 \alt \ln t \alt 6$, that
corresponds to $-2 \alt \ln (-\ln \Phi_{L}) \alt 0.75$, i.e. $0.87
\agt \Phi_{L} \agt 0.12$. For $\gamma = 15$, Fig. \ref{fig3}, the
value of $\beta$ in the intermediate time window is $\beta \simeq
0.567$, and the interval of the relaxation function described by it
is $0.92 \agt \Phi_{L} \agt 0.12$. Finally, for $\gamma =50$, Fig.
\ref{fig4}, it is $\beta \simeq 0.534$ for $0.95 \agt \Phi_{L} \agt
0.12$.

The kind of behavior of the response function has been often found
in the context of structural glasses, both in experiments and in
simple models, for the linear relaxation of the energy following a
temperature perturbation \cite{Sch86,Fr88,ByP96}. Then, this is
another indication of a possible conceptual connection between
structural glasses and dense granular systems. Moreover, the
simulation results for the model clearly indicate that the value of
the exponent $\beta$ characterizing most of the relaxation tends to
$1/2$ as the value of $\gamma$ increases. In simple models of
structural glasses, this value is associated to diffusive processes
\cite{ByP96,BPyR94}.

Although an explicit analytical solution of the model has not been
derived yet, it is possible to understand the diffusive origin of
the exponent $\beta=1/2$ as well as to get some information about
the characteristic relaxation time $\tau$. For $\gamma \gg 1$ ($X
\ll 1$), a typical metastable configuration of the system consists
of isolated holes separated a distance of the order of $\gamma$
(see, for instance, Eq.\ (\ref{2.15})). In this situation, the only
relevant process to decrease the volume is diffusion: holes move
through the system until two of them get together and one is
destroyed. In this picture, the characteristic relaxation time will
have the diffusion form  $\tau \propto \gamma^{2} =e^{2/X}$, i.e. an
Arrhenius type law with the molecular temperature replaced by  $X$.

\begin{figure}
\includegraphics[scale=0.45]{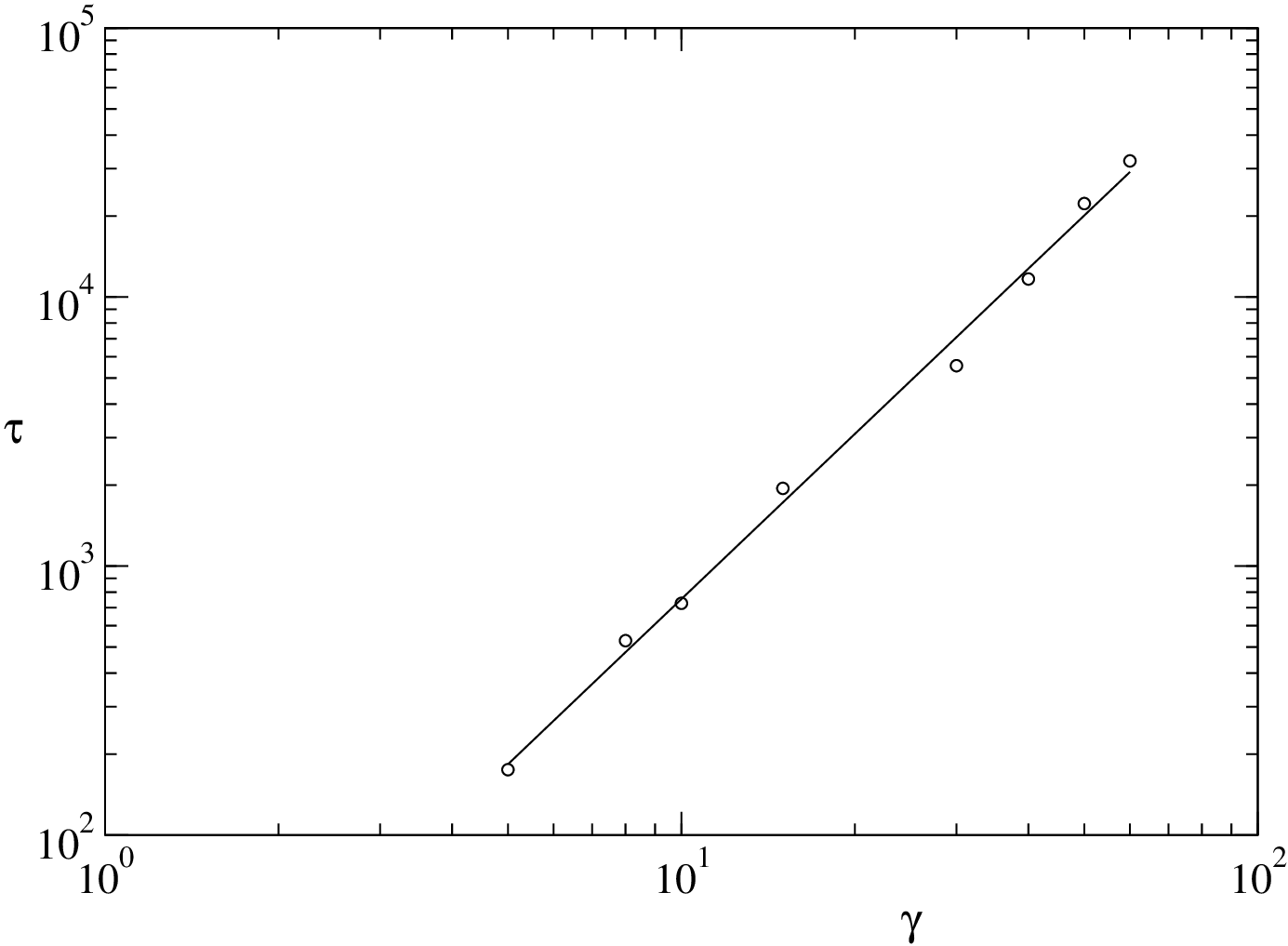}
\caption{Plot of the relaxation time $\tau$ appearing in the KWW
expression, Eq. \protect{(\ref{5.12})}, as a function of the rate
ratio $\gamma$. Time is measured in units proportional to the number
of taps. The symbols are from the Monte Carlo simulation data and
the solid line a linear fit of them. \label{fig5}}
\end{figure}

In Fig.\ \ref{fig5}, the Monte Carlo simulation results for $\tau$,
obtained by fitting the relevant intermediate time window to a KWW
function as discussed above, are plotted as a function of $\gamma$.
Note the logarithmic scale, so that the observed straight line
represents a power law behavior. The best fit, given by the solid
line, yields $\tau \approx 6.87 \gamma^{2.04}$ in good agreement
with the result following from the diffusive picture.

\section{Conclusion}
\label{s6} A simple model for compaction in granular media has been
used to investigate some static and dynamic properties of the steady
sate reached by the system in the long time limit. The simplicity of
the model allows to obtain analytical expressions for some of its
properties. On the other hand, it reproduces qualitatively well many
of the properties found in real granular systems. In the context of
static properties, this includes the Gaussian shape of the volume
fluctuations, the dependence of its variance on the average volume
and the number of particles, and the possibility of identifying the
compactivity as defined by Edwards.

Moreover, the scenario obtained from  the analysis of the model,
provides an useful tool to interpret the experimental results and
molecular dynamic simulations of more realistic models. This is
exemplified in the critical revision carried out in Sec. \ref{s4},
where the need of a clear differentiation between density
fluctuations at constant volume and at constant number of particles
shows up. Also, attention must be paid to distinguish between local
and global applications of the thermodynamic theory.

Another issue that is worth commenting is the possible existence of
negative values of the compactivity and/or the effective
temperature. If the mechanical statistical description is assumed as
the starting point of a thermodynamic description, there is no solid
reason to deny by principle this possibility. Even more, one is
tempted to conclude that it is highly probable, at least at a
theoretical level. Negatives values of the compactivity will show up
unavoidably if the number of metastable configurations is not a
monotonic increasing function of the volume. The possible volume
occupied by the tapped granular medium is limited by the loose and
random jamming packings. In both limits, the number of possible
metastable configurations can be expected to be quite reduced. The
requirement of maximum or minimum space between particles impose a
rather severe restriction on the allowed arrangement of the
particles. Therefore, it seems sensible to expect that the number of
configurations for intermediate values of the volume be larger than
at the extremes, presenting a maximum at some values. The same
qualitative reasoning applies to the effective or configurational
temperature changing the volume by the energy.

The question then is why negative compactivities (effective
temperatures) have been not identified in the experiments yet. There
are two main reasons for it. The first one is that only differences
of compactivities have been measured up to now, and it has often
been arbitrarily assumed that the compactivity diverges in the
random loose packing limit. The second reason is that negative
values of the compactivity, if they exist, would correspond to the
lowest density theoretically accesible region, and it can be very
hard to reach in practice. In any case, this is a point that
definitely deserves much more study.

In the last part of the paper, the linear relaxation of the model
has been considered. A slow relaxation, accurately represented  by a
KWW function with exponent $\beta$ tending to $1/2$ as the parameter
corresponding to the vibration intensity is decreased, has been
found. Moreover, the expression of the characteristic relaxation
time $\tau$ has Arrhenius form (with the consistent substitution of
the compactivity by the temperature).

The behavior of the linear response function for the model is fully
similar to what is observed in the relaxation of many structural
glasses and also in models of glassy relaxation. It must be noted
that the model discussed here was built to mimic the compaction
experiment and that, consistently, the relaxation of the density
starting from a loose configuration exhibits the characteristic
inverse logarithmic law \cite{ByP03a} found in experiments
\cite{KFLJyN95,NKBJyN98}. It seems interesting to experimentally
investigate whether linear response in tapped granular systems is
also slow and described by a KWW function. If that were the case, it
would be another indication of similarity between two apparently
different kind of systems, granular media and structural glasses,
reinforcing the link already found in other phenomena \cite{GyS06},
including the existence of hysteresis cycles \cite{NKBJyN98,PByS00}
and memory effects \cite{JTMyJ00,ByP02}.

\begin{acknowledgements}
This research was supported by the Ministerio de Educaci\'{o}n y
Cienc\'{\i}a (Spain) through Grant No. FIS2005-01398 (partially
financed by FEDER funds).
\end{acknowledgements}

\end{document}